\documentclass[twocolumn,longauthor]{aastex62}
\turnoffediting

\usepackage{amsmath}

\graphicspath{ {.} } 

\begin{document}

\title{A Scaling for Atmospheric Heat Redistribution on Tidally-Locked Rocky Planets.}

\correspondingauthor{Daniel D.B. Koll}
\email{dkoll@pku.edu.cn}

\author{Daniel D.B. Koll}
\affil{Department of Atmospheric and Oceanic Sciences, Peking University, Beijing, China}

\begin{abstract}
Atmospheric heat redistribution shapes the remote appearance of rocky exoplanets but there is currently no easy way to predict a planet's heat redistribution from its physical properties.
This paper proposes an analytical scaling theory for the heat redistribution on tidally-locked rocky exoplanets.
The main parameters of the scaling are a planet's equilibrium temperature, surface pressure, and broadband longwave optical thickness.
The scaling compares favorably against idealized general circulation model simulations of TRAPPIST-1b, GJ1132b, and LHS 3844b. For these planets heat redistribution generally becomes efficient, and a planet's observable thermal phase curve and secondary eclipse start to deviate significantly from that of a bare rock, once surface pressure exceeds $\mathcal{O}(1)$ bar.
The scaling additionally points to planetary scenarios for which heat transport can be notably more or less efficient, such as H$_2$ and CO atmospheres or hot lava ocean worlds.
The results thus bridge the gap between theory and imminent observations with the \textit{James Webb Space Telescope}. They can also be used to parameterize the effect of 3D atmospheric dynamics in 1D models, thereby improving the self-consistency of such models.
\end{abstract}


\section{Introduction}
\label{sec:intro}

Telescope observations have recently begun to probe the atmospheres of small, rocky, exoplanets around nearby stars \citep{demory2016,wit2018,diamond-lowe2018,kreidberg2019}. These observations will only get better in the near future with the launch of the \textit{James Webb Space Telescope} (\textit{JWST}) and the construction of extremely large ground-based telescopes.
The data that we will be able to obtain with these instruments promise insight into fundamental questions such as how rocky exoplanets form, how many of them host atmospheres, and whether any of them might be habitable.

To interpret these observations correctly, however, we need to understand the chemical and physical processes that shape the observable features of rocky exoplanets. These processes include radiative transfer and molecular absorption \citep{seager2000}, gas-phase chemistry \citep{moses2011,hu2014a}, clouds and hazes \citep{horst2018,moran2018}, as well as large-scale atmospheric dynamics \citep{joshi1997,merlis2010}.
Of these processes atmospheric dynamics has an outsized impact on thermal observations, because it determines a planet's global heat redistribution which sets the depth of the planet's secondary eclipse as well as the shape of the planet's thermal phase curve.

Unfortunately heat redistribution remains poorly represented in most models that are being used to match and interpret exoplanet observations.
The underlying reason is that these retrieval models have to be fast enough to be run $\sim 10^5-10^6$ times \citep{madhusudhan2009,line2013b}, which is required to comprehensively map out which atmospheric scenarios can, or cannot, fit an observed dataset.
As a consequence most retrieval models are idealized one-dimensional representations of a planet's atmosphere  \citep[e.g.,][]{tinetti2007,madhusudhan2009,benneke2012,line2013b,kempton2017a}.
By virtue of being 1D, these models cannot resolve 3D processes such as clouds and atmospheric heat redistribution, which has been shown to bias the results of atmospheric retrievals as well as model mean states \citep{line2016a,feng2016,fauchez2018}.

In principle 3D processes including atmospheric heat redistribution can be resolved by more complex models, such as general circulation models (GCMs). In practice 3D models are computationally far too costly to be used in place of 1D retrieval models. 
For example, the GCM used later in this paper requires about 24 hours of computation time on 8 processors. Even a dedicated supercomputer, running 32 GCM simulations in parallel for an entire month, would thus only be able to explore $\sim10^3$ parameter combinations, falling far short of being useful for standard retrieval techniques.

An important question is therefore how exoplanet retrieval models should represent the basic physics that determines observable quantities such as a planet's dayside emission or its thermal phase curve, while remaining computationally cheap.
Most 1D models represent the effects of heat redistribution by adjusting the planet's dayside energy budget, which can be written as \citep{burrows2014}
\begin{eqnarray}
  T_{day} & = & T_* \sqrt{\frac{R_*}{d}} (1-\alpha_B)^{1/4} f^{1/4}, \nonumber \\
          & = & (4 f)^{1/4} T_{eq}.
  \label{eqn:intro}
\end{eqnarray}
Here $T_{day}$ is the observed dayside brightness temperature\footnote{$T_{day}$ is hotter than the average dayside temperature because the observer viewing geometry is skewed towards the hot substellar point \citep{cowan2008}.}, $T_*$ is the stellar temperature, $R_*$ is the stellar radius, $d$ is the planet's semi-major axis, $\alpha_B$ is the
planet's bond albedo, and $f$ is the so-called heat redistribution
factor.
{It is useful to re-express $T_{day}$ in terms of the equilibrium temperature $T_{eq} = T_* \sqrt{R_*/(2d)} (1-\alpha_B)^{1/4}$, which is the temperature of an isothermal sphere with the same bond albedo and semi-major axis as the planet. The temperature $T_{eq}$ is closely related to the planet's characteristic fluxes, in particular $\sigma T_{eq}^4$, the flux received from the host star averaged across the entire planet, $2 \sigma T_{eq}^4$, the stellar flux averaged across the planet's dayside only, and $4 \sigma T_{eq}^4$, the incoming flux at the planet's substellar point.
The redistribution factor $f$ has to lie between $2/3$ for a planet without an atmosphere and $1/4$ for a planet that is extremely efficient at redistributing heat \citep{hansen2008a}, but currently there no easy way of expressing $f$ in terms of a planet's physical properties.

The goal of this paper is to develop a simple scaling theory for $f$.
The derivation is shown in Section \ref{sec:theory}. The results demonstrate that $f$ is sensitive to a planet's equilibrium temperature, surface pressure, and its broadband optical thickness, although the scaling also describes the dependency on other physical parameters.
In Section \ref{sec:gcm} the scaling is compared against GCM simulations of three nearby rocky planets: TRAPPIST-1b, GJ1132b, and LHS3844b \citep{gillon2016,delrez2018,berta-thompson2015,vanderspek2019}, which are among the highest-priority targets for upcoming thermal observations with \textit{JWST}.
The scaling successfully captures the dominant processes that determine atmospheric heat redistribution on these planets. The scaling
therefore lends insight into the atmospheric dynamics of tidally locked planets. 
Moreover, it can be used as a computationally efficient parameterization of large-scale dynamics in 1D models.
Because the derivation and GCM simulations assume idealized semi-grey radiative transfer, 
Section \ref{sec:computing_tau} considers how this work can be applied to real gases with non-grey absorption.
The results are discussed in Section \ref{sec:discussion}, and the conclusions are in Section \ref{sec:conclusion}.

\section{A scaling for atmospheric heat redistribution} \label{sec:theory}

\edit1{
This section derives a day-night heat redistribution scaling by combining two arguments. First, tidally locked planets are relatively slow rotators, so their atmospheric dynamics are to first order in weak-temperature-gradient (WTG) balance  \citep{pierrehumbert2011b,mills2013}.
Second, planetary atmospheres resemble heat engines, which means their day-night circulation strength and wind speeds can be approximately predicted using Carnot's theorem \citep{koll2016,koll2018}.
The combination of these two arguments allows us to analytically predict a planet's day-night heat redistribution, at least in the limit of thin atmospheres with inefficient heat redistribution. The prediction is then extended to arbitrary atmospheric thicknesses using an Ansatz. The choice of an Ansatz solution can only be justified retrospectively. It does appear useful, however, as the scaling's predictions compare favorably against a wide range of GCM simulations.
}

\edit1{The core idea of WTG balance is to reduce the atmosphere's full equation of energy conservation into a balance between radiative cooling and warming by adiabatic compression. We can expect this balance to hold for parcels of air that are undergoing large-scale descent, in particular the atmosphere on the nightside of a slowly-rotating tidally locked planet for which horizontal temperature gradients are small.}
Using the grey optical depth $\tau$ as the vertical coordinate, the balance can be written as \citep[see][]{koll2016}
\begin{eqnarray}
  \frac{c_p \omega}{g}\left( \frac{d T}{d\tau} - \beta \frac{T}{\tau}\right) & = &
                                                               \frac{d F}{d\tau}.
\label{eqn:wtg01}
\end{eqnarray}
Here $T(\tau)$ is the atmosphere's nightside vertical temperature profile, $F(\tau)$
is the net infrared flux, $\omega$ is the vertical velocity, $c_p$ is the atmosphere's specific heat capacity, $g$ is the acceleration of gravity, $\beta \equiv R/(c_p n_{LW})$ is
the dry adiabatic lapse rate in optical depth coordinates, $R$ is the
atmosphere's specific heat constant, and $n_{LW}=1$ if opacity is independent of
pressure (e.g., if molecular line widths are set by thermal broadening)
while $n_{LW}=2$ for pressure broadening. \edit1{The vertical velocity is defined in terms of a parcel's rate of pressure change, $\omega \equiv Dp/Dt$, and for sinking air $\omega>0$. The optical depth $\tau$ varies from zero at the top-of-atmosphere to $\tau_{LW}$ at the surface.}
If one evaluates $F$ at the top-of-atmosphere one obtains the nightside's flux of outgoing longwave radiation, $F_n\equiv F(\tau=0)$. \edit1{As long as internal heat fluxes (e.g., tidal heating) are negligible, this has to} equal the day-night heat transport by the atmosphere. \edit1{Therefore, if we can use Equation \ref{eqn:wtg01} to constrain $F_n$, we will also know how much heat the atmosphere is redistributing.}

\edit1{It is difficult to solve Equation \ref{eqn:wtg01} in general, because the infrared flux $F(\tau)$ and temperature profile $T(\tau)$ are closely coupled. However, previous work showed from simple energy-balance considerations that in the limit of a thin atmosphere the atmosphere becomes roughly isothermal and approaches the grey skin temperature, so $T(\tau)\approx T_{skin} = 2^{-1/4}  T_{eq}$ \citep{wordsworth2015}. The underlying reason is that an optically thin atmosphere still absorbs a trickle of radiation from the warm dayside surface which, as long as horizontal temperature gradients are small, allows the entire atmosphere to equilibrate at the temperature $T_{skin}$. A roughly isothermal atmosphere also significantly simplifies Equation \ref{eqn:wtg01},}
\begin{eqnarray}
  - \frac{c_p \omega}{g} \beta \frac{T_{skin}}{\tau} & \approx & \frac{d F}{d\tau}.
\label{eqn:wtg02}
\end{eqnarray}

\edit1{If we specify $\omega$, we can integrate Equation \ref{eqn:wtg02} to find the top-of-atmosphere flux $F_n$. In general one can expect $\omega = f(\tau) \times \omega_0(p_s,...)$, where $f(\tau)$ needs to go to zero at the top-of-atmosphere and at the surface to ensure that the atmospheric flow vanishes at those boundaries, while $\omega_0$ captures the still-to-be-determined dependence of the circulation on all other atmospheric parameters (such as surface pressure $p_s$). The detailed structure of $f(\tau)$ does not matter much, because it only ends up yielding a numerical constant. For example, assuming $f(\tau)=\sin(\pi \tau/\tau_{LW})$ and using the boundary condition that the nightside surface has to be in radiative equilibrium, $F(\tau_{LW})=0$, integration leads to:}
\begin{eqnarray}
  F(\tau_{LW}) - F_n & \approx & - \frac{c_p \omega_0}{g} \beta T_{skin} \int_0^{\tau_{LW}} \frac{\sin(\pi \tau/\tau_{LW})}{\tau} d\tau,\nonumber\\
  F_n & \approx & \frac{c_p \omega_0}{g} \beta T_{skin} \times 1.85,\nonumber\\
      & \approx & \frac{c_p \omega_0}{g} \beta T_{skin}
\label{eqn:wtg03}
\end{eqnarray}
\edit1{where we have dropped the numerical factor of order unity.}

\edit1{Next, we introduce Carnot's theorem to constrain the circulation strength $\omega_0$. In steady state, an atmospheric heat engine needs to balance the generation of work against frictional dissipation, so Carnot's theorem can be written as}
\begin{eqnarray}
    \rho_s C_d U^3 = \eta Q.
\label{eqn:carnot}
\end{eqnarray}
\edit1{Here the left-hand side expresses the rate of frictional dissipation, $\rho_s = p_s/(R T_s)$ is the air density near the surface, $C_d$ is a drag coefficient, $U$ is the near-surface wind speed, $\eta = (T_{hot}-T_{cold})/T_{hot}$ is the heat engine's efficiency, and $Q$ is the heat flux absorbed by the atmosphere. In the limit of a thin atmosphere the dayside surface is heated by stellar radiation, but can re-emit most of this energy directly back to space as longwave radiation. The remaining flux that is available to heat the atmosphere is therefore only $Q = (1-e^{-\tau_{LW}}) \times 2 \sigma T_{eq}^4\approx \tau_{LW} \times 2 \sigma T_{eq}^4$, }
\edit2{where $(1-e^{-\tau_{LW}})$ is the fraction of radiation not directly re-emitted to space according to Beer's law and $2 \sigma T_{eq}^4$ is the incoming stellar flux averaged across the planet's dayside. }
\edit1{If the areas of upwelling air on the dayside and downwelling air on the nightside are roughly equal, the vertical velocity on the nightside is then simply determined by mass balance, $U/L \approx \omega_0/p_s$, where $L$ is the circulation's horizontal scale. }
\edit2{Combining the expressions in this paragraph with Carnot's theorem (Eqn.~\ref{eqn:carnot}) yields the following vertical velocity scale,}
\begin{eqnarray}
    \omega_0 & = & \frac{p_s}{L}\left( \frac{\eta \tau_{LW} 2 \sigma T_{eq}^4}{\rho_s C_d} \right)^{1/3}.
\end{eqnarray}
\edit1{Plugging back into Equation \ref{eqn:wtg03}, the day-night heat redistribution of a thin atmosphere is thus}
\begin{eqnarray}
  F_n \approx \chi \frac{c_p \beta}{g L} \left( \frac{R p_s^2 \tau_{LW} \sigma T_{eq}^8}{C_d} \right)^{1/3}\nonumber,\\
  F_n \approx \chi \frac{c_p \beta}{g L} \left( \frac{R p_s^2 \tau_{LW} }{C_d \sigma^2 T_{eq}^4} \right)^{1/3} \sigma T_{eq}^4,
\label{eqn:Fn_thin}
\end{eqnarray}
\edit1{where we have absorbed all numerical constants as well as the heat engine efficiency $\eta$ into the factor $\chi$.}
\edit2{Because the values that enter $\chi$ are all roughly order unity, and because most of these values are further raised to the one-third power, one might expect that $\chi$ should also be of order one. In practice, however, we do not try to predict the value of $\chi$ from first principles. Instead we treat $\chi$ as the major source of uncertainty when comparing the scaling to numerical simulations (see below).}

\edit1{Finally, we seek to generalize the asymptotic solution for thin atmospheres in Equation \ref{eqn:Fn_thin}
using an Ansatz.} To do so note that Equation \ref{eqn:Fn_thin}
cannot be generally valid because it predicts
that $F_{n} \rightarrow \infty$ as $p_s$ or $\tau_{LW}$ become
large, whereas energy conservation requires that $F_{n} \rightarrow \sigma
T_{eq}^4$ for a thick atmosphere.
A general solution needs to reduce to the asymptotic solution
as $p_s$ and $\tau_{LW}$ become small, but tend towards $\sigma T_{eq}^4$ as
one or both parameters become large. \edit1{The} following expression
satisfies both requirements:
\begin{eqnarray}
  F_{n} & = & \frac{\chi \frac{c_p \beta}{g L} \left( \frac{R p_s^2 \tau_{LW} }{C_d \sigma^2 T_{eq}^4} \right)^{1/3}}{1 + \chi \frac{c_p \beta}{g L} \left( \frac{R p_s^2 \tau_{LW} }{C_d \sigma^2 T_{eq}^4} \right)^{1/3}} \sigma T_{eq}^4.
\label{eqn:Fn}
\end{eqnarray}
There is no guarantee that Equation \ref{eqn:Fn} will be exact
when \edit1{the factor preceding $\sigma T_{eq}^4$} is of order unity. Nevertheless, because \edit1{the expression} reduces to the correct limits for both a thin and a thick atmosphere, it is physically motivated.

To transform $F_n$ into a prediction for the dayside flux $F_d$ one can use the planet's energy budget, \edit1{$1/2 (F_d + F_n) = \sigma T_{eq}^4$, so $F_d = 2\sigma T_{eq}^4 - F_n$. However, this argument does not yet account for the observer's skewed viewing geometry.} At secondary eclipse the observer's view is weighted towards the hot substellar point, which directly faces the observer, and is less sensitive to colder regions that lie close to the
terminator, which tilt away from the observer \citep{cowan2008}.
The limiting expressions for the \edit1{observed dayside flux are given by Equation \ref{eqn:intro}, namely $F_{d,obs}=\sigma T_{day}^4=8/3 \times \sigma T_{eq}^4$ for a bare rock, and $F_{d,obs}=\sigma T_{eq}^4$ for a planet with uniform heat redistribution.}
\edit1{Since the scaling which interpolates between the bare-rock and uniform limits on the nightside should also apply on the dayside,} the observed dayside flux is thus
\begin{eqnarray}
  F_{d,obs} & = & \left( \frac{8}{3} - \frac{5}{3}\frac{\tau_{LW}^{1/3}
              \left(\frac{p_s}{1\text{bar}}\right)^{2/3} \left(\frac{T_{eq}}{600\text{K}}\right)^{-4/3}}{k + \tau_{LW}^{1/3}
              \left(\frac{p_s}{1\text{bar}}\right)^{2/3}\left(\frac{T_{eq}}{600\text{K}} \right)^{-4/3}} \right)\sigma T_{eq}^4.
\label{eqn:Fobs02}
\end{eqnarray}
Here $k=Lg/(\chi\beta c_p)\times (C_d \sigma^2/R)^{1/3} (1 \mathrm{bar})^{-2/3} (600 \mathrm{K})^{4/3}$ captures all planetary parameters other than the optical thickness, surface pressure, and equilibrium temperature. \edit1{Detailed evaluation below shows that, as long as $R$ and $c_p$ correspond to high mean-molecular-weight (MMW) atmospheres,} $k$ is \edit1{roughly} of order unity and \edit1{tends to vary little between common} planetary scenarios (e.g., radius and surface gravity tend to vary by less than a factor of two between different rocky planets). \edit1{We therefore separate out this parameter to underline the dominant dependency of heat redistribution on $p_s$, $\tau_{LW}$, and $T_{eq}$.}

To relate the main result back to the heat redistribution factor $f$ used in 1D models, one \edit1{simply has to divide the factor preceding $\sigma T_{eq}^4$ in Equation \ref{eqn:Fobs02} by four (see Eqn.~\ref{eqn:intro}):}
\begin{eqnarray}
  f & = & \frac{2}{3} - \frac{5}{12} \times
                        \frac{\tau_{LW}^{1/3} \left(\frac{p_s}{1
                        \text{bar}}\right)^{2/3}\left(\frac{T_{eq}}{600\text{K}}\right)^{-4/3}}{k
                        +\tau_{LW}^{1/3} \left(\frac{p_s}{1
                        \text{bar}}\right)^{2/3}\left(\frac{T_{eq}}{600\text{K}}\right)^{-4/3}}.
\label{eqn:final_result}
\end{eqnarray}
As expected, Equation \ref{eqn:final_result} recovers the no-redistribution limit $f \rightarrow 2/3$ as $\tau_{LW},p_s \rightarrow 0$, and the uniform-redistribution limit $f \rightarrow 1/4$ as $\tau_{LW},p_s \rightarrow \infty$.

\section{Testing theory with GCM Simulations}
\label{sec:gcm}

\subsection{Numerical setup}

\begin{deluxetable}{lllll}
\tablecaption{Planetary parameters for the GCM simulations.}
\tablecolumns{5}
\tablehead{
\colhead{} & \colhead{Radius} & \colhead{Period} &\colhead{Surf.~gravity} & \colhead{T$_{eq}$\tablenotemark{*}} \\
\colhead{} & \colhead{(R$_\Earth$)} & \colhead{(days)} &\colhead{(m/s$^2$)} & \colhead{(K)}
}
\startdata
TRAPPIST-1b & 1.12 & 1.51 & 7.95 & 391 \\
GJ1132b & 1.16 & 1.63 & 11.8 & 578 \\
LHS3844b & 1.32 & 0.46 & 12.9\tablenotemark{**} & 805
\enddata
\label{tab:parameters}
\tablenotetext{*}{Equilibrium temperature, for uniform heat redistribution and zero albedo.}
\tablenotetext{**}{~Assuming 2.3 $M_{\Earth}$, based on \citet{chen2017a}.}
\end{deluxetable}

To test the analytical \edit2{scaling}
\edit1{this paper uses} the Flexible Model System (FMS) general circulation model (GCM) with dry thermodynamics. FMS is a widely-used model which has previously been applied to the atmospheres of Earth \citep{frierson2006}, Jupiter \citep{liu2011a},
hot Jupiters \citep{heng2011a}, tidally locked terrestrial planets
\citep{merlis2010,mills2013,koll2015,koll2016,hammond2017}, as well as rapidly rotating
terrestrial planets \citep{kaspi2015}.
Consistent with dry thermodynamics, \edit1{the model does not include} the radiative impact of clouds or photochemical hazes. \edit1{Section \ref{sec:discussion} discusses the potential shortcomings} of these modeling assumptions.

The FMS version \edit1{used here} simulates the atmosphere's full 
large-scale dynamics coupled to semi-grey (shortwave versus longwave) radiative
transfer. Convection is parameterized as instantaneous dry convective adjustment. Near-surface turbulence is parametrized using a standard Monin-Obukhov scheme which
self-consistently computes the depth of the boundary layer
as well as turbulent diffusion of heat and momentum.  The surface is
represented by a ``slab layer'', that is a single layer
with uniform temperature and fixed layer depth. Simulations are all
tidally locked and orbits are assumed to be circular so that the
stellar flux is fixed in space and constant in time.

\edit1{Atmospheric shortwave absorption and surface albedo are set to zero}, so all stellar energy is \edit1{deposited} at the planet's surface.
Similar to \citet{frierson2006} \deleted{I assume} the longwave optical
thickness depends quadratically on pressure in the lower atmosphere,
which represents the effects of pressure broadening, while it depends
linearly on pressure in the upper atmosphere, which represents the effect
of thermal broadening and helps ensure that the stratosphere
equilibrates within reasonable run times.

\edit1{The horizontal resolution is T42, which is} equivalent to about $64 \times 128$ points in latitude and longitude, \edit1{while the vertical resolution is 30 grid points}. As is standard in GCMs, FMS includes horizontal hyperdiffusion which acts as a kinetic energy filter at the smallest length scales resolved by the model. Although such a filter can be potentially problematic in modeling gas giant atmospheres, because the physical processes that lead to frictional dissipation in gas giants are often not explicitly modeled and thus hyperdiffusion acts as a stand-in for unresolved physics \citep{koll2018}, this issue is less important for rocky planets where friction from the solid surface is captured by the Monin-Obukhov boundary layer scheme.

Three rocky planets that orbit nearby M-dwarfs \edit1{are simulated}:
TRAPPIST-1b, GJ1132b, and LHS3844b. The planetary parameters are shown in Table \ref{tab:parameters}.
\edit1{Observations of these planets disfavor H$_2$-rich atmospheres \citep{wit2018,diamond-lowe2018,kreidberg2019}, so the specific gas constant $R$ and the specific heat capacity $c_p$ are set to their values for N$_2$ to represent a generic high MMW atmosphere.}
The three planets span a wide range of equilibrium temperatures, which is one of the dominant parameters
in the \edit2{scaling} 
(Section \ref{sec:theory}). In addition, the three planets span different rotational regimes. LHS3844b has an orbital period of about 11h, whereas TRAPPIST-1b and GJ1132b have orbital periods of 1.5 and 1.6 days, which translates into a nondimensional Rossby deformation radius of $a/L_{Ro} \approx 3$ for LHS3844b versus $a/L_{Ro} \approx 2$ for TRAPPIST-1b and GJ1132b. \edit1{Here $a/L_{Ro} =\sqrt{2\Omega a/c_{wave}}$, where $a$ is the planet radius, $\Omega$ is the rotation rate, and $c_{wave}=\sqrt{R/c_p}\sqrt{R T_{eq}}$ is a characteristic gravity wave speed.} Given that the theoretical scaling does not account for the effect of planetary rotation, comparison between the three planets thus also provides a check on whether the \edit2{scaling} 
is robust to changes in planetary rotation.

Consistent with the CFL criterion, \edit1{simulations} are
more likely to crash at higher stellar fluxes. \edit1{The simulations} therefore use a numerical diffusion coefficient of $2.31\times10^5$ s$^{-1}$
(damping time of half a day) for TRAPPIST-1b and GJ1132b, and between $9.26\times10^4$ to $2.31\times10^4$ s$^{-1}$ for LHS3844b. In some cases \edit1{the timestep also needs to be reduced at low surface pressure or high optical thickness}. The default timestep is 120 s for TRAPPIST-1b, 60 s for GJ1132b, and 20 s for LHS3844b.

\edit1{The simulations explore the} effect on the atmosphere's heat redistribution of varying atmospheric surface pressure $p_s$ and optical thickness $\tau_{LW}$.
\edit1{To consider a wide combination of both parameters, surface pressure is varied between 0.01 bar and 100 bar, and the longwave optical thickness is set to 0.1, 1, and 5.}
Motivated by the observation that solar system atmospheres have broadband optical thicknesses of about
1-10 at 1 bar \citep{robinson2014b}, \edit1{the simulations also include} a scenario in
which \edit1{optical thickness $\tau_{LW}$ increases linearly} with surface
pressure $p_s$ using the relation $\tau_{LW}= p_s/(\text{1 bar})$.

\begin{figure*}
\includegraphics[width=\linewidth,clip]{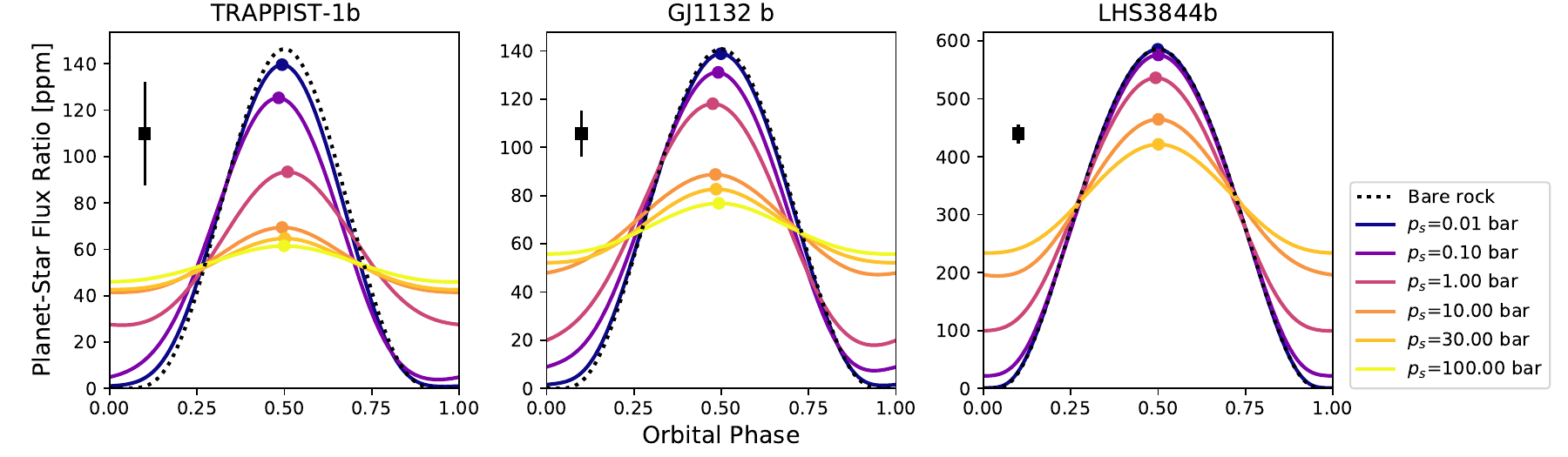}
\caption{Dayside emission starts
  to significantly deviate from that of a bare rock once surface pressure exceeds $\mathcal{O}(1)$ bar. Lines show simulated phase curves, and circles
  show hot spots (i.e., a planet's maximum thermal emission).
  Hot spot offsets are negligible and all hot spots remain close to secondary eclipse at orbital phase = 0.5. Black error bars show a lower bound for
  the 1$\sigma$ photometric precision possible with \textit{JWST}'s MIRI instrument (photon noise integrated \edit1{over} the duration of each planet's transit), so the planets' day-night thermal variations should be observable for all three targets.}
\label{fig:02}
\end{figure*}

\subsection{GCM results}

\begin{figure*}
\includegraphics[width=\linewidth,clip]{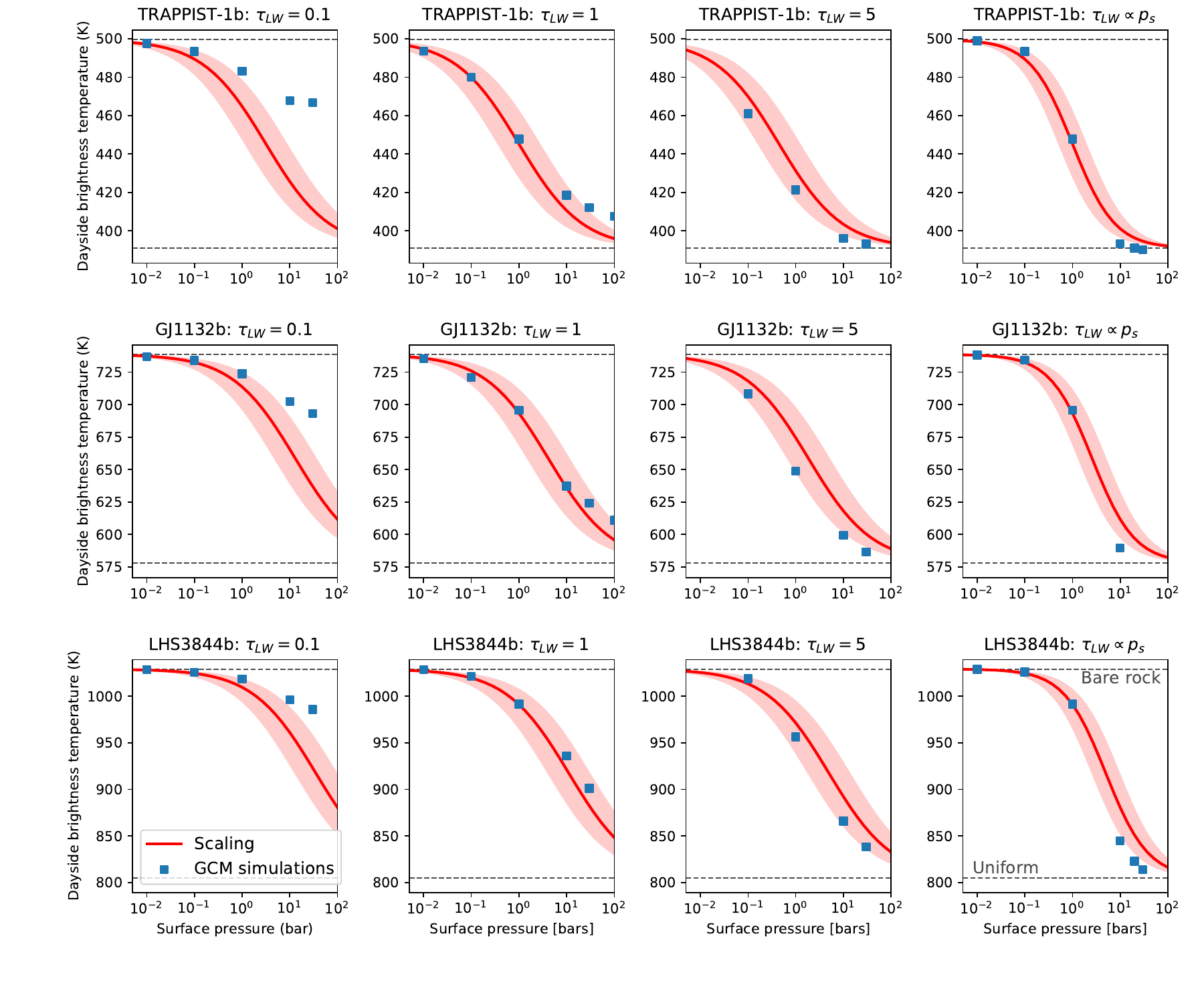}
\caption{The theoretical scaling captures how a planet's dayside thermal emission depends on atmospheric and planetary parameters.
  Dots are GCM simulations, the red line is the theoretical scaling, and the light red envelope indicates a representative factor of two uncertainty in the heat engine efficiency parameter that enters the \edit2{scaling}. 
  Horizontal dashed lines show the limiting cases of a bare rock (zero
  heat redistribution) and uniform redistribution.
  \label{fig:03}}
\end{figure*}

Figure \ref{fig:02} shows that thermal phase curves start to significantly deviate from a bare rock once surface pressure exceeds about 1 bar, in agreement with the analytical scaling.
Shown are results for simulations with constant optical thickness, $\tau_{LW}=1$. 
\edit1{The thermal phase curves are computed} across the \textit{JWST} MIRI bandpass (5-12$\mu$m).
To convert a planet's simulated thermal emission to a planet-star flux ratio \edit1{the stellar spectrum is additionally assumed to be a blackbody}.
At surface pressures of 0.1 bar or less phase curves are effectively indistinguishable from a bare rock.
This is particularly the case once \textit{JWST}'s instrumental precision \edit1{is included}, which cannot be any better than the photon noise limit but could optimistically be comparable to it (Fig.~\ref{fig:02}).
Conversely, at surface pressures above 10 bar phase curves become increasingly uniform. This matches the theoretical prediction \edit1{expressed in Equation~\ref{eqn:Fobs02}, namely that as long as $k$ is of order unity, the transition between a bare rock and a uniform planet also happens at $\mathcal{O}(1)$ bar. } Moreover, for a fixed surface pressure, LHS3844b has the least efficient redistribution while TRAPPIST-1b has the most efficient redistribution. This matches the expectation that, at fixed surface pressure \edit1{and optical thickness}, redistribution \edit1{becomes less efficient on hotter planets.}

Figure \ref{fig:03} compares the theoretical prediction against the dayside brightness temperatures that an observer would see at secondary eclipse.
To evaluate $k$ in Equation \ref{eqn:final_result} \edit1{we assume a high MMW atmosphere with $(R,c_p)=(R_{N_2},c_{p,N_2})$, $C_d=1.9\times 10^{-3}$, and $L=a$.}
\edit1{To constrain $\chi$ note that previous work showed some GCM simulations of tidally locked planets can indeed approach the ideal limit described by Carnot's theorem; more typical simulations, however, find wind speeds that fall a factor of two to four below those predicted for an ideal heat engine \citep{koll2016}. The underlying reason is that other atmospheric processes, such as small-scale convection, also redistribute heat but do so without generating large-scale kinetic energy. Based on comparison with the GCM experiments this paper adopts $\chi=0.2$, but the value should be treated as uncertain by at least a factor of two since $\chi$ accounts not only for the atmosphere's thermodynamic efficiency but also the circulation geometry.}

\edit1{Adopting the above values, $k$ only depends weakly on a planet's exact properties, with $k=1.2$ for TRAPPIST-1b, $k=1.9$ for GJ1132b, and $k=2.3$ for LHS3844b. At least across the range of planets considered here the factors determining $k$ (e.g., planetary radius) are thus of secondary importance for a planet's heat redistribution compared to variations in $p_s$, $\tau_{LW}$, and $T_{eq}$. The notable exception to this rule would be low MMW atmospheres.
Because $k \propto 1/(c_p R^{1/3}) \sim \mathrm{MMW}^{4/3}$, the scaling predicts that planets with H$_2$-dominated atmospheres will have significantly lower $k$, and thus more efficient heat redistribution, than planets with CO$_2$-, N$_2$-, or H$_2$O-dominated atmospheres. This result matches previous theoretical work on exoplanet heat transport which also found that heat redistribution is significantly more efficient in low MMW atmospheres \citep{menou2012a,heng2012c,zhang2017c}.}

Figure \ref{fig:03} shows that the scaling correctly captures the main variation in dayside thermal emission across several orders of magnitude variation in surface pressure, and more than one order of magnitude variation in optical thickness. 
The scaling is not perfect, however.
\edit1{For example, the scaling tends to slightly underpredict thermal emission at low optical thickness, while it overpredicts thermal emission at high optical thickness. This suggests that day-night redistribution is somewhat more sensitive to changes in $\tau_{LW}$ than is implied by the scaling, which in the optically thin limit predicts $F_n \propto \tau_{LW}^{1/3}$ (Eqn.~\ref{eqn:Fn_thin}).
There are also significant differences between scaling and simulations for atmospheres with high surface pressure and low optical thickness, in particular TRAPPIST-1b with $\tau_{LW}=0.1$ and $p_s>10$ bar. Although the combination of such high surface pressure and low optical thickness is probably unrealistic (see Section \ref{sec:computing_tau}), the scaling becomes inaccurate in this regime. One potential reason is that at high surface pressure convective heat fluxes should become important. Convection implies that the atmosphere ceases being isothermal and instead tends towards a vertically adiabatic profile, which would reduce the efficiency with which adiabatic compression can heat the nightside (see Equation \ref{eqn:wtg01}).}

\begin{figure*}
\includegraphics[width=\linewidth,clip]{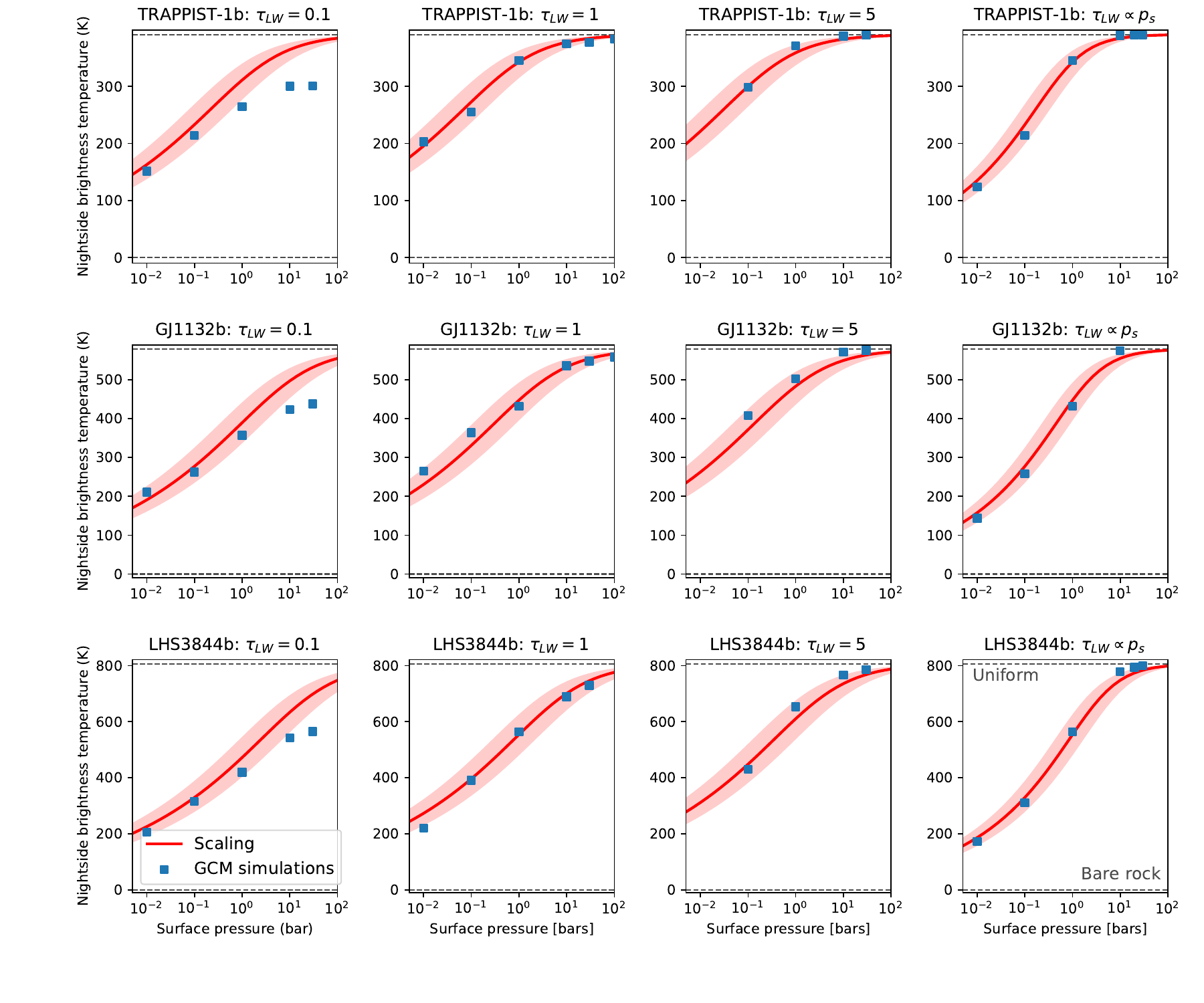}
\caption{The theoretical scaling captures how a planet's nightside thermal emission depends on atmospheric and planetary parameters.
  Dots are GCM simulations, the red line is the theoretical scaling, and the light red envelope indicates a representative factor of two uncertainty in the heat engine
  efficiency parameter that enters the \edit2{scaling}.
  Horizontal dashed lines show the limiting cases of a bare rock (zero
  heat redistribution) and uniform redistribution.
  \label{fig:04}}
\end{figure*}

Nevertheless, the majority of simulations in Figure \ref{fig:03} fall inside the red envelope around the theoretical scaling. This indicates that the differences between \edit2{scaling} 
and simulations can be largely explained by a factor-of-two inaccuracy in the derivation. One potential source of inaccuracy is the inefficiency factor $\chi$, which \edit1{the scaling assumes to be} constant but which could vary if the atmosphere's heat engine efficiency shifted in response to changes in surface pressure or optical thickness. 
Another potential source of inaccuracy is the effect of planetary rotation\edit1{, which is not captured by the scaling}. For a given planet the rotation rate is constant, so one might naively expect a roughly constant mismatch between \edit2{scaling}
and simulations. A planet's rotation interacts with the large-scale circulation, however, to induce spatially varying flow and temperature patterns which could shift with changes in pressure or optical thickness and thus create a varying mismatch between \edit2{scaling}
and simulations.

Figure \ref{fig:04} shows that the scaling also captures the main trends in nightside brightness temperatures.
This is not surprising - given that the \edit2{scaling}
largely captures the dayside's thermal emission, energy conservation implies that the \edit2{scaling} 
should match the nightside thermal emission with similar accuracy. As on the dayside, the majority of nightside brightness temperatures agree with the scaling to within a factor of two change in $k$.
This means the \edit2{scaling} 
is able to \edit1{roughly capture how changes in atmospheric and planetary properties affect} a planet's dayside and nightside thermal emission, which is important because one or both of these quantities can be directly measured via secondary eclipse and thermal phase curve observations.

\section{Assigning an equivalent grey optical thickness}
\label{sec:computing_tau}

Up to now \edit1{the derivation has assumed} grey radiative transfer, but real gases have an optical thickness that varies strongly as a function of wavelength. How can one assign an equivalent optical thickness $\tau_{LW}$ to atmospheres with realistic compositions? \edit1{We} tested Rosseland and Planck mean opacities but found that they provide a poor fit.
Instead \edit1{this section will focus on} the degree to which an atmosphere diminishes the surface's thermal emission, and thus reduces a planet's observable surface emission below that of a bare rock.

For an atmosphere with spectrally varying absorption, the planet's top-of-atmosphere thermal flux is equal to
\begin{eqnarray}
  F & = & \pi \int B_\lambda(T_s) e^{-\tau_\lambda} d\lambda +
          \left(\mathrm{atm.~emission}\right).
\end{eqnarray}
Here $\lambda$ is wavelength, $T_s$ is the surface temperature, the first term is the surface's blackbody emission attenuated by the overlying atmosphere, and the second term is the atmosphere's emission. For comparison, with a grey absorber the top-of-atmosphere thermal flux is equal to
\begin{eqnarray}
  F & = & \sigma T_s^4 e^{-\tau_{LW}} +
          \left(\mathrm{atm.~emission}\right),
\end{eqnarray}
where $\tau_{LW}$ is now the atmosphere's grey optical thickness.

By setting the two surface terms equal to each other, one can define the equivalent grey optical thickness for any atmospheric composition as
\begin{eqnarray}
  \tau_{LW} & \equiv & -\ln\left[  \frac{\int e^{-\tau_\lambda}
                  B_\lambda(T_s)d\lambda}{\int
                  B_\lambda(T_s)d\lambda} \right].
\end{eqnarray}
This equivalent grey optical thickness has the desirable property that it exactly matches the extent to which an atmospheric column attenuates the surface's thermal flux. It is therefore most appropriate for the thin-atmosphere limit, in which most of a planet's thermal emission originates at the surface, and it correctly captures how the gradual increase of atmospheric mass then reduces the planet's surface thermal emission below that of a bare rock.

Figure \ref{fig:grey} shows $\tau_{LW}$ for four representative atmospheric compositions, namely pure CO$_2$, H$_2$O, O$_3$, and CO. To calculate $\tau_{LW}$ a 1D radiative transfer model with line-by-line spectral resolution \edit1{was used} \citep{koll2018a}\footnote{Available at \url{https://github.com/ddbkoll/PyRADS}.}. The surface temperature is set to the equilibrium temperature of LHS3844b and the overlying atmosphere follows a dry adiabat.
\edit1{All molecular opacities are taken from the HITRAN2016 database \citep{gordon2017}. Collision-induced absorption (CIA) is included for H$_2$O and CO$_2$ using the fits from \citet{pierrehumbert2010}.}
\edit1{The HITRAN2016 database does not include weak lines, which become important opacity sources at high temperatures. Similarly, we are not aware of CO-CO or O$_3$-O$_3$ CIA data, even though CIA should become important at high surface pressures. The optical thicknesses shown in Figure \ref{fig:grey} should therefore be treated as a lower bound.}

\begin{figure}
\includegraphics[width=\linewidth,clip]{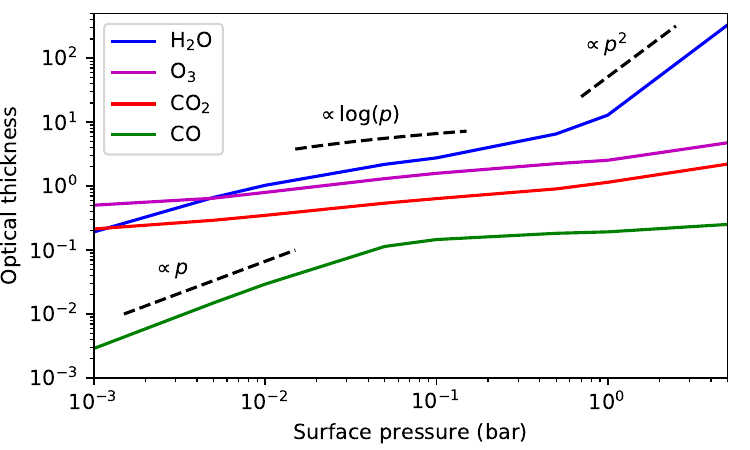}
\caption{Equivalent grey optical thickness for actual gases, \edit1{assuming a surface temperature representative of LHS3844b.} \deleted{I assume an atmospheric column filled with a single gas, a dry adiabatic temperature structure, and a surface temperature representative of LHS3844b.} H$_2$O is a strong absorber and is effectively optically thick above $\sim10^{-2}$ bar. O$_3$ and CO$_2$ become optically thick above $\sim 1$ bar. The main exception is a poor absorber like CO, which is still optically thin at 1 bar.}
\label{fig:grey}
\end{figure}

H$_2$O is a very strong absorber, with 10$^{-2}$ bar of H$_2$O enough to make the atmosphere optically thick. Next follow CO$_2$ and O$_3$, which become optically thick once surface pressure exceeds about 1 bar. Finally, CO is the poorest absorber out of these gases, and is optically thin even at a surface pressure of 1 bar.
The rate at which $\tau_{LW}$ increases with pressure can be roughly understood as follows. At low surface pressure $\tau_{LW}$ is dominated by the behavior of individual molecular absorption lines, so absorption is proportional to pressure \citep[see strong and weak line limits in][]{pierrehumbert2010}. H$_2$O and CO both follow this expectation below about $10^{-2}$ bar. 
At high surface pressures absorption becomes dominated by CIA, which scales with pressure squared. H$_2$O follows this expectation above $1$ bar.
Finally, for CO$_2$ and O$_3$, as well as CO above 0.1 bar, $\tau_{LW}$ is set by the closing of a molecule's window regions and thus a molecule's overall band shape. For a molecular band for which \edit1{the absorption cross-section} decays roughly exponentially away from the band center, this implies a logarithmic scaling with pressure \citep{pierrehumbert2010,koll2018a}.

Figure \ref{fig:grey} shows that for an atmosphere with a moderately strong greenhouse effect, such as CO$_2$, the equivalent optical thickness of a 1 bar atmosphere is of order unity. Combined with \edit1{the heat transport scaling (Eqn.~\ref{eqn:final_result}), and assuming a high MMW atmosphere for which $k$ is of order unity,} this implies that \edit1{a} typical surface pressure above which the thermal phase curve of a rocky exoplanet with $T_{eq}\sim 600$ K starts to deviate from that of a bare rock is also about 1 bar.
\edit1{However, the scaling underlines that there are clear exceptions to this rule-of-thumb: in addition to low MMW atmospheres, for which $k$ becomes much smaller than unity, atmospheres with a particularly strong or weak greenhouse effect will also exhibit a more or less efficient heat transport (e.g., steam or pure CO atmospheres).}

\section{Discussion}
\label{sec:discussion}

The scaling \edit1{proposed} here provides first-order insight into the processes that determine the day-night thermal contrasts of tidally locked, rocky exoplanets. The scaling builds on the numerical results of \citet{koll2015} and the theory of \citet{koll2016}, and translates them into observable quantities that can be measured in the near future with \textit{JWST} via secondary eclipse and phase curve measurements.

Although the scaling relies on grey radiation for its derivation, it appears to agree with previous GCM studies that used comprehensive radiative transfer.
\citet{selsis2011} found that the transition between large day-night contrasts and uniform emission on short-period rocky planets \edit1{with CO$_2$ atmospheres} occurs somewhere between 0.1 and 1 bars. The scaling here agrees with their results and predicts a transition at 0.4 bar, assuming their planetary parameters and $\tau_{LW}\sim 1$. Similarly, \citet{yang2013} and \citet{wolf2019} simulated \edit1{habitable} planets with 1 bar \edit1{N$_2$-H$_2$O atmospheres}. \deleted{and full H$_2$O thermodynamics (condensation and clouds).} These studies found thermal phase curves that are fairly uniform relative to a bare rock, and thus also support the scaling here which predicts a transition at about 0.2 bar. \edit1{Note that the last two studies simulated habitable planets with a full hydrologic cycle, which allows the atmosphere to transport more heat than predicted by the scaling (see below). Habitable planets might thus exhibit efficient heat redistribution at surface pressures even below 0.2 bar.}
%

%
%

\edit1{The} \edit2{scaling}
\edit1{still contains some fundamental uncertainties. In addition to the inefficiency factor $\chi$, the derivation relies on an Ansatz to generalize from thin atmospheres to arbitrary surface pressures. } \edit2{Neither does the derivation account for the effects of planetary rotation}.
\edit1{Figures \ref{fig:03} and \ref{fig:04} show that the basic} \edit2{approach} \edit1{is broadly justified, but also suggest that the match with the GCM simulations tends to be better at low surface pressures than at high pressures.}
\edit1{Future analytical work could therefore focus on the} \edit2{limit of extremely thick atmospheres, and consider a wider range of planetary rotation rates, to improve the scaling in those regimes.}

\edit1{Another} outstanding issue is that the thermal emission of rocky exoplanets can be affected by additional atmospheric and planetary physics not included here.
\edit1{Broadly speaking, we expect that the scaling} should overpredict a planet's day-night thermal contrast.
First, the \edit2{scaling}
assumes dry thermodynamics, but an atmosphere with condensation can additionally transport heat via latent heat transport \citep{ding2018}. The derivation here does not capture this process, so it should tend to overpredict day-night thermal contrasts in atmospheres in which condensation starts to become important.
\edit1{Second,} some exoplanets might be able to sustain a surface ocean, in particular planets inside their host star's habitable zone. In addition to \edit1{supporting a hydrologic cycle and latent heat transport}, an ocean can additionally transport energy via its \edit1{own} fluid motions which will again reduce the planet's day-night thermal contrast \citep{yang2019}.

\edit1{Finally, the scaling proposed here does not account for the radiative effect of clouds, but it is unclear whether clouds should systematically increase or decrease a planet's day-night emission contrast. For clouds that are linked to atmospheric convection, such as H$_2$O clouds on habitable planets, clouds will occur preferentially on a planet's dayside and thus reduce the planet's day-night emission contrast \citep{yang2013}. However, on extremely hot planets with large day-night temperature contrasts, such as 55 Cnc e, clouds can also form by advection of Na or SiO vapor to the planet's cold nightside \citep{hammond2017}. In this case clouds would preferentially amplify the planet's day-night emission contrast.}

\section{Conclusion}
\label{sec:conclusion}

\edit1{This paper proposes} an analytical scaling \edit1{for atmospheric heat redistribution} on tidally locked rocky exoplanets.
\deleted{The scaling compares favorably against a large set of idealized GCM simulations.}
\edit1{The surface pressure at which a planet's broadband secondary eclipse and thermal phase curve transitions from inefficient to efficient heat redistribution depends on the atmosphere's bulk composition, its radiative properties, and the planet's equilibrium temperature.}

\edit1{For ``warm'' planets like TRAPPIST-1b, GJ1214b, and LHS3844b, a typical threshold pressure is $\mathcal{O}(1)$ bar, but there are notable exceptions. H$_2$-dominated atmospheres redistribute heat much more efficiently than high-MMW atmospheres.} \edit2{Heat redistribution also depends on optical thickness, with optically thick H$_2$O atmospheres better at redistributing heat than optically thin CO atmospheres.} \edit1{Finally, \textit{JWST} will also be able to observe thermal emission from hot ``lava ocean'' worlds, such as 55 Cancri e and K2-141b. Becaus heat redistribution is less efficient on hotter planets, the scaling predicts that these planets require atmospheres thicker than $\mathcal{O}(1)$ bar surface pressure to exhibit significant heat redistribution.}

In addition to improving our understanding of \edit1{atmospheric heat transport}, the scaling is useful in the context of \deleted{modeling and} interpreting thermal observations of rocky exoplanets. For favorable targets, \textit{JWST} will be able to measure broadband thermal fluxes as well as emission spectra \citep[e.g., ][]{greene2016,morley2017,batalha2018,kempton2018}.
To make sense of these measurements, models are needed that properly account for the physical processes which shape the observable features of rocky exoplanets. The scaling derived here offers a way of parameterizing the large-scale atmospheric heat transport in 1D models, and was used as such in a number of related studies \citep{kreidberg2019,koll2019a,mansfield2019,malik2019}. Doing so is attractive because parameterization is many orders of magnitude faster than explicitly simulating an atmosphere's large-scale fluid dynamics.

\acknowledgments

This work was supported by a James McDonnell Foundation postdoctoral fellowship.
\edit1{The paper was improved by constructive comments from an anonymous reviewer.}

\bibliography{main}

\end{document}